\documentclass[aps,prl,preprint,groupedaddress]{revtex4-2}
\usepackage{amsmath}
\usepackage{amssymb}
\usepackage{graphicx}
\usepackage{float}
\usepackage{natbib}
\usepackage[utf8]{inputenc}
\usepackage{soul}

\begin{document}

\title{Conservation law for angular momentum based on optical field derivatives: Analysis of optical spin-orbit conversion}

\author{Shun Hashiyada}
\email{shun.hashiyada@es.hokudai.ac.jp}
\affiliation{Laboratory of Nanosystem Optical Manipulation, Department of Photonics and Optical Science, Research Institute for Electronic Science, Hokkaido University, Nishi 10, Kita 21, Kita-ku, Sapporo, Hokkaido 001-0021, Japan}

\author{Yoshito Y. Tanaka}
\email{ytanaka@es.hokudai.ac.jp}
\affiliation{Laboratory of Nanosystem Optical Manipulation, Department of Photonics and Optical Science, Research Institute for Electronic Science, Hokkaido University, Nishi 10, Kita 21, Kita-ku, Sapporo, Hokkaido 001-0021, Japan}

\date{\today}

\begin{abstract}
We present a theoretical framework for analyzing the loss of optical angular momentum (AM), including spin (SAM) and orbital (OAM) components, in light-matter interactions.  Conventional SAM and OAM conservation laws rely on transverse field components, neglecting longitudinal fields and limiting applicability to vacuum.  Our approach defines optical AM using time derivatives of the electric and magnetic fields, yielding a gauge-invariant formulation that includes both transverse and longitudinal components and explicitly incorporates charge and current densities.  This enables a more complete description of AM dissipation in materials. We apply this framework to analyze spin-orbit conversion (SOC) in two scenarios: scattering of circularly polarized (CP) beams by a gold nanoparticle and focusing of CP and linearly polarized optical vortex beams by a lens.  The results show that SOC depends on particle size and polarization, with notable OAM loss in larger particles and CP beam focusing.  This framework enables the evaluation of previously overlooked SAM and OAM losses, providing a powerful tool for studying systems in which the analysis of AM losses is intrinsically important, such as chiral materials, as well as for designing photonic devices and exploring light-matter interactions at the nanoscale.
\end{abstract}

\maketitle

\section{ I. INTRODUCTION}
The conserved quantities of optical fields, including energy, optical chirality (or Lipkin’s Zilch), linear momentum, and angular momentum (AM), are altered when light interacts with matter through processes such as optical absorption and scattering \cite{ref1, ref2, ref3, ref4, ref5, ref6, ref7}. Optical AM is lost when transferred to the material. This transfer not only generates optical torque, inducing the rotational motion of the material itself \cite{ref8, ref9, ref10, ref11, ref12, ref13, ref14, ref15, ref16, ref17, ref18, ref19, ref20}, but also drives internal elementary excitations with AM \cite{ref21, ref22, ref23, ref24, ref25, ref26, ref27, ref28, ref29, ref30, ref31, ref32, ref33, ref34, ref35, ref36}, including chiral plasmons and phonons. Chiral materials, such as helical structures, respond differently to light with positive and negative AM, a property known as circular (or helical) dichroism \cite{ref32, ref33, ref34, ref35, ref36}. These may result from lifted degeneracy between excitations with opposite AM.  Thus, optical AM loss reveals valuable information about the structures and properties of the material.

Light carries two distinct forms of AM: spin AM (SAM) and orbital AM (OAM). Optical SAM is associated with the intrinsic rotation of the electric and magnetic field vectors in circularly polarized (CP) light, where the field traces a helical path as the light propagates \cite{ref8, ref9}. On the other hand, optical OAM arises from the helical wavefront structure in optical vortices (OVs) \cite{ref37}. When CP OVs, carrying both optical SAM and OAM, interact with absorbing particles, the particles exhibit both spinning and revolving due to optical torque \cite{ref18}. While this behavior suggests a simple link between SAM and spinning, and OAM and revolving, the dynamics are more complex due to optical spin-orbit conversion (SOC) \cite{ref38, ref39, ref40, ref41, ref42}. SOC occurs when CP light is tightly focused or scattered, converting SAM into OAM and generating OVs. This complicates the analysis of SAM and OAM contributions to the optical torque. For instance, nanoscale CP optical fields generated by focusing lenses \cite{ref19} or properly designed plasmonic nanostructures \cite{ref20} can induce orbital motion in nanoparticles, though experimental identification of SOC from focusing, scattering, or both remains difficult. SOC has also been observed in metasurfaces \cite{refadd1}, where structured light fields interact with nanostructures to couple SAM and OAM.  In evanescent waves \cite{refadd2, refadd3}, spin-momentum locking or spin-orbit interaction leads to a strong correlation between the propagation direction and the polarization state. Conceptualizing OAM in the optical near field remains challenging, making its role in sub-wavelength light-matter interactions hard to interpret.

Theoretically, optical torque arises from the transfers of total optical AM and can be described by its conservation law, with density:
\begin{equation}
    \mathbf{J} = \mathbf{r} \times \left[ \varepsilon_0 (\mathbf{E} \times \mathbf{B}) \right],
\end{equation}
where $\mathbf{r}$ is the position vector, $\varepsilon_0$ is the permittivity of free space, and $\mathbf{E}(\mathbf{r},t)$ and $\mathbf{B}(\mathbf{r},t)$ are the time-dependent electric and magnetic fields, respectively. Local conservation law or continuity equation for total AM can be expressed as $ \frac{\partial \mathbf{J}}{\partial t} + \nabla \cdot \mathbf{M} = -\boldsymbol{\tau}$, where $\mathbf{M} = \mathbf{r} \times \left[ -\varepsilon_0 (\mathbf{E} \otimes \mathbf{E}) - \varepsilon_0 c^2 (\mathbf{B} \otimes \mathbf{B}) + \frac{\varepsilon_0}{2} (\mathbf{E} \cdot \mathbf{E} + c^2 \mathbf{B} \cdot \mathbf{B}) \mathbf{I} \right]$ is the flux density (where $c$ is the speed of light in free space), and $\boldsymbol{\tau} = \mathbf{r} \times \mathbf{F}_\text{Lorenz} = \mathbf{r} \times \left( \rho \mathbf{E} + \mathbf{j} \times \mathbf{B} \right)$ is the loss density, corresponding to the optical torque density. Materials are regarded as a distribution of charges with density $\rho(\mathbf{r}, t)$ and currents with density $\mathbf{j}(\mathbf{r}, t)$. Since total AM is a conserved quantity, the effective (time-averaged) loss $\mathbf{T}$ due to light-matter interaction can be quantitatively evaluated by calculating the time-averaged total AM flux on a surface enclosing the material as $\mathbf{T} = \int_V \langle \boldsymbol{\tau} \rangle dV = -\oint_S \langle \mathbf{M} \rangle \cdot \mathbf{n} dS$ \cite{ref7}, where time-averaged quantities are denoted by $\langle \cdot \rangle$, $\mathbf{n}$ is the outward normal unit vector to the surface, and $S$ is an arbitrary closed surface surrounding the material. However, this method alone does not allow for distinguishing whether the observed optical torque arises from the loss of SAM or OAM.  To address this, it is necessary to separately derive the conservation laws for SAM and OAM and evaluate their respective losses independently.

Conventionally, the separation of SAM and OAM from hl{total} AM has been performed using the transverse components of the electromagnetic field and the vector potential, ensuring gauge invariance \cite{ref43}. However, this approach has a fundamental limitation: it neglects the contributions of the longitudinal components, which are essential in the presence of materials. As a result, the conservation laws for SAM and OAM derived within this framework assume a material system that generates only transverse currents \cite{ref44, ref45, ref46}. However, such an assumption is not physically valid, as real materials inevitably generate both transverse and longitudinal components of the field and the potential. Consequently, the formulation can only be applied in a vacuum, where charge and current densities are zero \cite{ref47}. To address these limitations, in this Letter, we introduce a novel framework in which the total AM is constructed from the time derivatives of the electric and magnetic fields, with density:
\begin{equation}
    \mathbf{J}' = \mathbf{r} \times \left[ \frac{\varepsilon_0}{\omega^2} \left( \frac{\partial \mathbf{E}}{\partial t} \times \frac{\partial \mathbf{B}}{\partial t} \right) \right],
    \label{eq:Jofd}
\end{equation}
where $\omega$ is the angular frequency. Within this framework, SAM and OAM are inherently gauge-invariant because they are expressed purely in terms of the full electromagnetic field, without reliance on the vector potential.  Furthermore, since the formulation includes both transverse and longitudinal components, it accounts for material-induced interactions, unlike conventional descriptions.  As a result, Maxwell’s equations for systems containing materials can be used to derive the conservation laws for SAM and OAM. This allows conservation laws to explicitly include charge and current densities and naturally incorporate SAM and OAM dissipation. While it has long been known that conservation laws hold not only for the original physical quantities composed of the fields but also for those composed of the field derivatives \cite{ref45,ref46,ref48,ref51,ref52,ref53}, the importance of considering field-derivative-based physical quantities has not been recognized so far. This is because, for monochromatic fields, the time-averaged form of conservation laws for field-derivative-based physical quantities often leads to the same conclusions as those based on conventional formulations. Importantly, there is no theoretical or experimental basis for determining a single "correct" form of optical AM, and thus both conventional AM and the AM defined in this work should be regarded as valid conserved quantities. In this study, we apply this field-derivative approach to conservation laws for SAM and OAM in systems containing materials, thus allowing us to account for previously overlooked losses of SAM and OAM.  This enables an explanation of SAM and OAM losses where SOC is significant.

\section{II. THEORETICAL FRAMEWORK}
We here first derive the continuity (conservation) equation for the total AM of time-derivative fields in systems containing materials.  By taking the time derivative of $\mathbf{J}'$ and using Maxwell’s equations, we get $\frac{\partial \mathbf{J}'}{\partial t} + \nabla \cdot \mathbf{M}' = -\boldsymbol{\tau}'$, where
\begin{equation}
    \mathbf{M}' = \mathbf{r} \times \left[ \frac{\varepsilon_0}{\omega^2} \left\{ - \frac{\partial \mathbf{E}}{\partial t} \otimes \frac{\partial \mathbf{E}}{\partial t} - c^2 \frac{\partial \mathbf{B}}{\partial t} \otimes \frac{\partial \mathbf{B}}{\partial t} + \frac{1}{2} \left( \frac{\partial \mathbf{E}}{\partial t} \cdot \frac{\partial \mathbf{E}}{\partial t} + c^2 \frac{\partial \mathbf{B}}{\partial t} \cdot \frac{\partial \mathbf{B}}{\partial t} \right) \mathbf{I} \right\} \right],
    \label{eq:Mofd}
\end{equation}
is the flux density, and
\begin{equation}
    \boldsymbol{\tau}' = \mathbf{r} \times \left[ \frac{1}{\omega^2} \left( \frac{\partial \rho}{\partial t} \frac{\partial \mathbf{E}}{\partial t} + \frac{\partial \mathbf{j}}{\partial t} \times \frac{\partial \mathbf{B}}{\partial t} \right) \right]
    \label{eq:Tofd}
\end{equation}
is the loss density. Assuming the time harmonicity of the fields $\mathbf{E}(\mathbf{r},t) = \Re[\tilde{\mathbf{E}}(\mathbf{r}) e^{-i\omega t}]$, $\mathbf{B}(\mathbf{r},t) = \Re[\tilde{\mathbf{B}}(\mathbf{r}) e^{-i\omega t}]$, $\mathbf{j}(\mathbf{r},t) = \Re[\tilde{\mathbf{j}}(\mathbf{r}) e^{-i\omega t}]$, and $\rho(\mathbf{r},t) = \Re[\tilde{\rho}(\mathbf{r}) e^{-i\omega t}]$, the time-averaged form of the total AM density $\langle \mathbf{J}' \rangle$, the flux density $\langle \mathbf{M}' \rangle$, and the loss density $\langle \boldsymbol{\tau}' \rangle$ correspond to those obtained from the conventional conservation law ($\langle \mathbf{J}' \rangle = \langle \mathbf{J} \rangle$, $\langle \mathbf{M}' \rangle = \langle \mathbf{M} \rangle$, and $\langle \boldsymbol{\tau}' \rangle = \langle \boldsymbol{\tau} \rangle$). Thus, for monochromatic fields, the time-averaged conservation law for the total AM of time-derivative fields matches the conventional one.

We next separate the total AM density $\mathbf{J}'$ into SAM density $\mathbf{J}_{\text{spin}}'$ and OAM density $\mathbf{J}_{\text{orbit}}'$ ($\mathbf{J}' = \mathbf{J}_{\text{spin}}' + \mathbf{J}_{\text{orbit}}'$). By using a simple vector identity to rewrite Eq. \ref{eq:Jofd}, we get the SAM density
\begin{equation}
    \mathbf{J}_{\text{spin}}' = \mathbf{r} \times \left[ \frac{\varepsilon_0}{\omega^2} \left\{ \left( \frac{\partial \mathbf{E}}{\partial t} \cdot \nabla \right) \mathbf{E}  \right\} \right],
    \label{eq:Jspin}
\end{equation}
and the OAM density
\begin{equation}
    \mathbf{J}_{\text{orbit}}' = \mathbf{r} \times \left[ - \frac{\varepsilon_0}{\omega^2} \left\{ \frac{\partial \mathbf{E}}{\partial t} \cdot (\nabla) \mathbf{E} \right\} \right],
     \label{eq:Jorbit}
\end{equation}
and we adopt the notation $\mathbf{X} \cdot (\mathbf{Y}) \mathbf{Z}$ for any quantities $\mathbf{X}$, $\mathbf{Y}$, and $\mathbf{Z}$. To confirm the physical meaning of $\mathbf{J}_{\text{spin}}'$ and $\mathbf{J}_{\text{orbit}}'$, we calculated the effective SAM and OAM by integrating the time-averaged quantities $\langle \mathbf{J}_{\text{spin}}' \rangle$ in Eq.\ref{eq:Jspin} and $\langle \mathbf{J}_{\text{orbit}}' \rangle$ in Eq.\ref{eq:Jorbit} in the xy-plane perpendicular to the direction of light propagation (z) for a CP Gaussian beam with $\tilde{E}_x^G = E_0 \exp \left\{ -\frac{x^2 + y^2}{w_0^2} + i(kz - \omega t) \right\}$ and $\tilde{E}_y^G = is \tilde{E}_x^G$ ($s = \pm 1$), and for a linearly (x-) polarized (LP) OV beam with $tilde{E}_x^{OV} = \tilde{E}_x^G \left( \sqrt{2} \cdot \frac{\sqrt{x^2 + y^2}}{w_0} \right)^{|l|} \exp (i l \phi)$ and $\tilde{E}_y^{OV} = 0$ ($l = 0, \pm 1, \pm 2, \dots$), where $E_0$, $w_0$, and $\phi$ are amplitude, beam waist radius, and azimuth angle in the beam cross section, respectively. For the CP Gaussian beam, the effective SAM is a finite value, and its sign depends on the handedness of the CP beam (the sign of $s$), whereas the effective OAM is null. On the other hand, for the LP OV beam, the effective SAM is null, while the effective OAM is a finite value, and its sign depends on the handedness of the OV beam (the sign of $l$). These confirm $\mathbf{J}_{\text{spin}}'$ and $\mathbf{J}_{\text{orbit}}'$ represent SAM and OAM. SAM density $\mathbf{J}_{\text{spin}}'$ in Eq.\ref{eq:Jspin} can be separated into the position-independent and -dependent parts as $\mathbf{J}_{\text{spin}}' = \frac{1}{\omega^2} \left\{ \mathbf{F} + \mathbf{G}(\mathbf{r}) \right\}$, where $\mathbf{F} = {\varepsilon_0} \left( \mathbf{E} \times \frac{\partial \mathbf{E}}{\partial t}  \right)$ and $\mathbf{G}(\mathbf{r}) = {\varepsilon_0} \left\{ \left( \frac{\partial \mathbf{E}}{\partial t} \cdot \nabla \right)(\mathbf{r} \times \mathbf{E}) \right\}$, respectively. The position-independent part $\mathbf{F}$ is known as the flux density of electric optical chirality (Lipkin’s Zilch) and can be regarded as the intrinsic (origin-independent) SAM density \cite{ref45, ref46, ref49}. In contrast, the position-dependent part $\mathbf{G}(\mathbf{r})$ can be regarded as the extrinsic (origin-dependent) SAM density.  Separating $\mathbf{J}_{\text{orbit}}'$ into position-dependent and -independent parts is mathematically difficult.

To compare the continuities of SAM and OAM densities of time-derivative fields, we take the time derivative of $\mathbf{J}_{\text{spin}}'$ and $\mathbf{J}_{\text{orbit}}'$, and we get $\frac{\partial \mathbf{J}_{\text{spin}}'}{\partial t} + \nabla \cdot \mathbf{M}_{\text{spin}}' = -\boldsymbol{\tau}_{\text{spin}}'$, and $\frac{\partial \mathbf{J}_{\text{orbit}}'}{\partial t} + \nabla \cdot \mathbf{M}_{\text{orbit}}' = -\boldsymbol{\tau}_{\text{orbit}}'$, respectively. The SAM flux density is
\begin{align}
    \mathbf{M}_{\text{spin}}' &= \frac{\varepsilon_0 c^2}{\omega^2} \left[ \frac{\partial \mathbf{B}}{\partial t} \otimes \mathbf{E} + \left\{\mathbf{E} \otimes \frac{\partial \mathbf{B}}{\partial t} \right. - \left( \mathbf{E} \cdot \frac{\partial \mathbf{B}}{\partial t}  \right) \mathbf{I}\right\}  - \frac{\partial}{\partial t} \left\{ \frac{1}{c^2} \frac{\partial \mathbf{E}}{\partial t} \otimes (\mathbf{r} \times \mathbf{E}) \right\} \Bigg],
    \label{eq:Mspin}
\end{align}
and the OAM flux density is
\begin{align}
\left( \mathbf{M}_{\text{orbit}}' \right)_{ij}
&= \frac{\varepsilon_0 c^2}{\omega^2} \Bigg[
  -\frac{\partial B_i}{\partial t} E_j \notag \\
&\quad + \epsilon_{ikl} r_k \Bigg\{
    \epsilon_{jmn} \left( \frac{\partial B_m}{\partial t} \, \frac{\partial E_n}{\partial r_l} \right)
    + \delta_{jl} \, \frac{1}{2} \left(
      \frac{1}{c^2} \frac{\partial \mathbf{E}}{\partial t} \cdot \frac{\partial \mathbf{E}}{\partial t}
      - \frac{\partial \mathbf{B}}{\partial t} \cdot \frac{\partial \mathbf{B}}{\partial t}
    \right)
  \Bigg\}
\Bigg]
\label{eq:Morbit}
\end{align}
where $\epsilon_{ikl}$ is the Levi-Civita symbol. The SAM loss density is
\begin{equation}
	\boldsymbol{\tau}_{\text{spin}}' = \frac{1}{\omega^2} \left[ -c^2 \rho \frac{\partial \mathbf{B}}{\partial t}  -  \frac{\partial \mathbf{j}}{\partial t} \times \mathbf{E} + \frac{\partial}{\partial t} \left\{  \frac{\partial \rho}{\partial t} (\mathbf{r} \times \mathbf{E}) \right\} \right],
	\label{eq:Tspin}
\end{equation}
and the OAM loss density is
\begin{equation}
	\boldsymbol{\tau}_{\text{orbit}}' = \frac{1}{\omega^2} \left[ c^2 \rho \frac{\partial \mathbf{B}}{\partial t} - \frac{\partial \mathbf{j}}{\partial t} \cdot (\mathbf{r} \times \nabla) \mathbf{E} \right].
	\label{eq:Torbit}
\end{equation}
To our knowledge, the loss densities for SAM in Eq. \ref{eq:Tspin} and OAM in Eq. \ref{eq:Torbit}, including charge and current densities of materials, are derived here for the first time. The first terms on the right-hand sides in Eqs. \ref{eq:Tspin} and \ref{eq:Torbit} are common terms with opposite signs. When the SAM and OAM losses are summed, these terms cancel out and do not contribute to the loss of total AM. This suggests that these terms represent a SOC where the SAM is converted to OAM and vice versa. SOC terms also appear in the flux densities in the first term on the right-hand sides in Eqs. \ref{eq:Mspin} and \ref{eq:Morbit}, in agreement with earlier findings \cite{ref47}.

In the continuity of SAM, the last terms on the right-hand sides in Eqs. \ref{eq:Mspin} and \ref{eq:Tspin} are time derivative forms, originating from the position-dependent part $\mathbf{G}(\mathbf{r})$, and thus, for monochromatic fields, the loss for $\mathbf{G}(\mathbf{r})$ vanishes after time-averaging. Therefore, only the position-independent part $\mathbf{F}$ contributes to the effective SAM loss, consistent with past results \cite{ref49}.

As discussed above, the first terms on the right-hand sides in Eqs. \ref{eq:Tspin} and \ref{eq:Torbit} are the SOC terms, thus the remaining (second) terms correspond to the losses due to the transfers of SAM and OAM from light to matter, respectively.

It is important to emphasize that in any electromagnetic response theory, light is described in terms of fields or potentials, while material systems are represented by charge and current densities, as dictated by Maxwell’s equations \cite{refadd5}. Our formulation follows standard theory and applies to various light-matter interactions, including atoms, molecules, and elementary excitations such as excitons. If a quantum mechanical treatment of the material system is required, this can be incorporated by solving charge and current densities within a quantum framework.  Thus, the conservation laws derived in this work remain valid even in the presence of quantum effects in the material system.

\section{III. ANALYSIS OF OPTICAL SPIN-ORBIT CONVERSION}
\subsection{A. Scattering of circularly polarized Gaussian beam by a single gold nanosphere}

\begin{figure}[H]
    \centering
    \includegraphics[width=0.5972\linewidth]{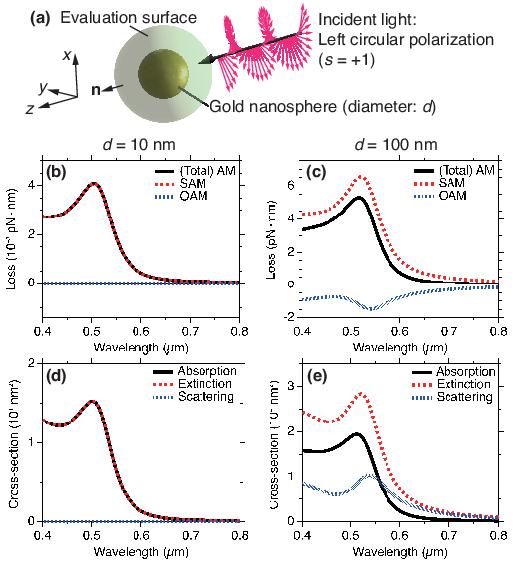}
    \caption{Analysis of optical spin-orbit conversion (SOC) in light scattering. (a) The geometric model of a single gold nanosphere in the air illuminated by left circularly polarized light of 1 mW/$\mu$m$^2$. (b,c) Wavelength dependence of losses of total optical angular momentum (AM), spin AM (SAM), and orbital AM (OAM) for spheres with diameters of 10 nm (b) and 100 nm (c). (d,e) Wavelength dependence of cross-sections of absorption, extinction, and scattering for spheres with diameters of 10 nm (d) and 100 nm (e).}
    \label{fig:soc_scattering}
\end{figure}

We applied our field-derivative-based framework to investigate SOC in the scattering of CP plane waves (Gaussian beam in the limit $w_0 \to \infty$) by gold nanoparticles (Fig. \ref{fig:soc_scattering}(a)). By calculating the time-averaged total AM flux density at the closed surface surrounding the sphere, the total AM loss in the propagation direction (+z) of the incident light can be obtained \cite{ref10} as
\begin{equation}
    T_z = \pm \frac{2 \pi \varepsilon_0 c^3 E_0^2}{\omega^3} \sum_{l}(2l + 1) \left\{ \text{Re}(a_l + b_l) - \left( |a_l|^2 + |b_l|^2 \right) \right\},
    \label{eq:Tstotal}
\end{equation}
where the sign corresponds to that of the SAM for the incident CP light, and $a_l$ and $b_l$ are the complex scattering coefficients of the $l$th-order electric and magnetic multipolar modes, respectively. Dipole and quadrupole modes correspond to $l = 1$ and $l = 2$, respectively. The loss of total AM is proportional to that of electromagnetic energy, which relates to the absorption cross-section \cite{ref50}. The first and second terms on the right-hand side of Eq. \ref{eq:Tstotal} originate from the extinction and the scattering, respectively.

The loss of SAM and OAM are evaluated by calculating $\mathbf{T}_{\text{spin}} = \int_V \langle \boldsymbol{\tau}_{\text{spin}}' \rangle dV = -\oint_S \langle \mathbf{M}_{\text{spin}}' \rangle \cdot \mathbf{n} dS$ and $\mathbf{T}_{\text{orbit}} = \int_V \langle \boldsymbol{\tau}_{\text{orbit}}' \rangle dV = -\oint_S \langle \mathbf{M}_{\text{orbit}}' \rangle \cdot \mathbf{n} dS$, respectively.  The SAM loss can be expressed as the difference between the loss of total AM in Eq. \ref{eq:Tstotal} and that of OAM ($T_z^{\text{spin}} = T_z - T_z^{\text{orbit}}$), where
\begin{multline}
    T_z^{\text{orbit}} = \pm \frac{2 \pi \varepsilon_0 c^3 E_0^2}{\omega^3} \sum_{l}(2l + 1) \Bigg[ \left( \frac{1}{l(l+1)} - 1 \right) \left( |a_l|^2 + |b_l|^2 \right) \\
    + \frac{2l(l+2)}{(l+1)(2l+1)} \text{Re} \left( a_l^* b_{l+1} + b_l^* a_{l+1} \right) \Bigg].
    \label{eq:Tsorbit}
\end{multline}
The optical OAM loss in Eq. \ref{eq:Tsorbit} depends only on quadratic terms of $a_l$ and $b_l$, indicating that only the scattered fields contribute to the SOC in this system. The first and second terms on the right-hand side of Eq. \ref{eq:Tsorbit} represent the SOC induced by the scattering from a pure single electric or magnetic mode and by the interference between the scattering of electric and magnetic modes with the same parity symmetry, respectively. The coefficient of the first term in Eq. \ref{eq:Tsorbit} is half of the total AM scattering loss in Eq. \ref{eq:Tstotal} for the dipolar mode ($l = 1$) and equal to that for sufficiently high-order modes ($l \gg 1$). Therefore, OVs carrying OAM are efficiently generated from the higher-order excitation modes of the material during the light scattering process.

When the incident left CP light is scattered by a single gold nanosphere with a diameter of 10 nm in air, positive SAM losses and negligible OAM losses are observed (Fig. \ref{fig:soc_scattering}(b)). The spectra for the losses of SAM and total AM nearly overlap.  This suggests that no SOC occurs, and the SAM of the incident left CP light is directly transferred to the material.  However, as the particle size increases to 100 nm, significant negative OAM losses are detected (Fig. \ref{fig:soc_scattering}(c)), indicating the generation of left OVs carrying a positive OAM. Thus, SOC strengthens with particle size.  Larger particles have greater scattering cross-sections and a higher probability of excitation of higher-order modes (Figs. \ref{fig:soc_scattering}(d) and \ref{fig:soc_scattering}(e)), leading to greater SAM-to-OAM conversion. Optical SAM, being a localized property, is less influenced by smaller particle sizes, whereas optical OAM, with its global nature, becomes more significant as particle size increases. These findings highlight the role of particle geometry in SOC and offer insights into particle interactions with CP light.

\subsection{B. Focusing of circularly polarized Gaussian beams and linearly polarized optical vortex beams by an aplanatic lens}

\begin{figure}[H]
    \centering
    \includegraphics[width=0.5972\linewidth]{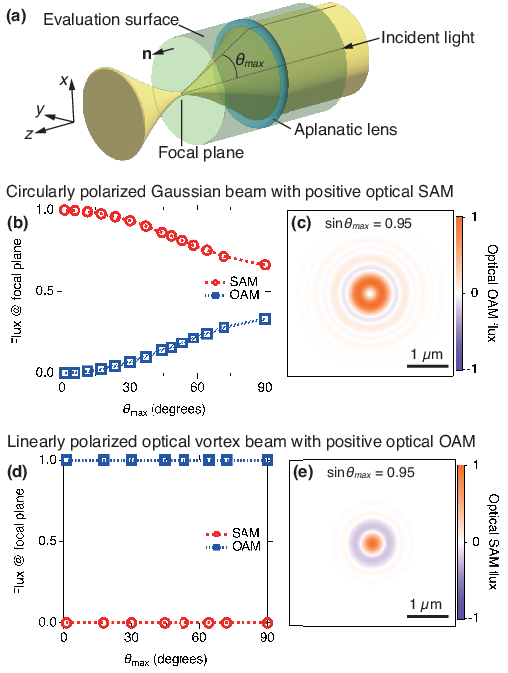}
    \caption{Analysis of optical spin-orbit conversion (SOC) in light focusing. (a) The geometric model of the aplanatic system. (b,d) Maximum focusing angle ($\theta_{\text{max}}$) dependence of the fluxes at the focal plane of optical spin angular momentum (SAM) and orbital angular momentum (OAM) for the incident circularly polarized Gaussian beam (b) and linearly polarized optical vortex beam (d) with a wavelength of 800 nm. The fluxes were normalized by the integrated total optical angular momentum (SAM + OAM) flux of the incident light. (c,e) The flux density maps of OAM (c) and SAM (e) at the focal plane under tightly focusing conditions with $\sin \theta_{\text{max}} = 0.95$. The flux densities were normalized by the maximum values in the images.}
    \label{fig:soc_focusing}
\end{figure}

In addition to light scattering, we analyzed SOC in the focusing of CP Gaussian beams by an aplanatic lens (Fig. \ref{fig:soc_focusing}(a)). Under weak focusing (small maximum focusing angle ($\theta_{\text{max}}$)), the normalized SAM flux remained unity and OAM flux was absent (Fig. \ref{fig:soc_focusing}(b)). However, as $\theta_{\text{max}}$ increased, SAM flux was gradually converted into OAM flux, evidenced by the corresponding decrease in SAM flux and increase in OAM flux. For the spatial distributions of the OAM flux in the focal plane under tight focusing conditions ($\sin \theta_{\text{max}} = 0.95$) (Fig. \ref{fig:soc_focusing}(c)), positive OAM flux became dominant, and even after spatial integration, a finite OAM flux remains, confirming the occurrence of SOC during the focusing of CP light by a lens. The extremal value of OAM flux originating from SOC reaches 1/3 of the total AM of the incident light, consistent with \cite{ref41}.

Next, we turned our attention to LP OV beams, which carry OAM but no SAM. Our analysis revealed that the OAM flux of the beam remained constant at unity, regardless of the focusing conditions, and no SAM flux was generated (Fig. \ref{fig:soc_focusing}(d)). At the focal plane, local SAM flux with opposite signs appeared (Fig. \ref{fig:soc_focusing}(e)). Upon spatial integration, these positive and negative local SAM flux contributions canceled out, resulting in a net zero SAM flux. This indicates that no SOC occurred in the case of focusing LP OV beams by a lens.

The absence of SOC in LP OV beams contrasts with CP Gaussian beam behavior, highlighting the crucial role of polarization in SOC. The interplay of SAM and OAM underscores how focusing influences SOC and optical torque. These findings offer practical insights into using beam manipulation to control SOC, which is critical for applications such as optical trapping and advanced photonic device development.

While previous SOC studies have primarily focused on simple optical systems where diffraction-based methods are sufficient, extending this analysis to more complex material structures, such as plasmonic nanostructures and metamaterials, and to more complex electromagnetic fields, such as evanescent waves and optical skyrmions \cite{refadd4}, presents significant challenges.  In such cases, analytical diffraction solutions become intractable, making it difficult to separate and quantify SAM and OAM losses.  The framework presented in this work provides a systematic approach to evaluating SAM and OAM dissipation in arbitrary material and electromagnetic field systems, enabling the quantitative characterization of SOC beyond the limitations of conventional methods.

\section{IV. CONCLUSION}
By introducing the concept of total optical AM based on time derivatives of the electric and magnetic fields, we proposed a method to analyze optical SAM and OAM losses in light-matter interactions. This framework extends SAM and OAM conservation laws to material systems by explicitly incorporating charge and current densities. It enables separate evaluation of optical SAM and OAM losses, improving understanding of AM dissipation. We applied this framework to study SOC in both light scattering by gold nanoparticles and light focusing by lenses. Our results confirmed that SOC occurs in CP light, with significant OAM loss depending on particle geometry and focusing angle, whereas no SOC occurs in focusing LP OVs. These findings demonstrate how polarization influences SOC and optical AM dissipation in light-matter interactions.

Unlike previous SOC studies that have primarily focused on simple optical systems where diffraction-based methods suffice, this framework extends SOC analysis to more complex material structures, such as plasmonic nanostructures and metamaterials, and to more complex electromagnetic fields, such as evanescent waves and optical skyrmions \cite{refadd4}, where spin-momentum coupling and near-field interactions play a crucial role in AM transfer. In such cases, conventional diffraction-based approaches become analytically intractable, making it difficult to separate and quantify optical SAM and OAM losses. The systematic formulation developed in this work overcomes these limitations by enabling the quantitative characterization of SOC in arbitrary material and electromagnetic field systems.  Our approach opens avenues to study SOC in complex systems, such as chiral materials, and explore optical torque, spin-orbit interactions, and AM conservation in nanophotonics. Future work will apply this framework to the analysis of circular (or helical) dichroism of chiral mate-rials and spin-orbit interaction systems to further elucidate nanoscale light-matter interactions.

\begin{acknowledgments}
We acknowledge fruitful discussions with Prof. Jun-ichiro Kishine, Prof. Stephen M. Barnett, and Dr. Frances Crimin. This work was supported by Grants-in-Aid for Scientific Research (KAKENHI) (Nos. JP24H00424, JP22H05132 in Transformative Research Areas (A) “Chiral materials science pioneered by the helicity of light” to Y.Y.T., and Nos. JP23K04669, JP21K14594 to S.H.) from the Japan Society for the Promotion of Science (JSPS), and JST FOREST Program (No. JPMJFR213O to Y.Y.T.).
\end{acknowledgments}

\bibliography{refs}

\begin{thebibliography}{58}%
\makeatletter
\providecommand \@ifxundefined [1]{%
 \@ifx{#1\undefined}
}%
\providecommand \@ifnum [1]{%
 \ifnum #1\expandafter \@firstoftwo
 \else \expandafter \@secondoftwo
 \fi
}%
\providecommand \@ifx [1]{%
 \ifx #1\expandafter \@firstoftwo
 \else \expandafter \@secondoftwo
 \fi
}%
\providecommand \natexlab [1]{#1}%
\providecommand \enquote  [1]{``#1''}%
\providecommand \bibnamefont  [1]{#1}%
\providecommand \bibfnamefont [1]{#1}%
\providecommand \citenamefont [1]{#1}%
\providecommand \href@noop [0]{\@secondoftwo}%
\providecommand \href [0]{\begingroup \@sanitize@url \@href}%
\providecommand \@href[1]{\@@startlink{#1}\@@href}%
\providecommand \@@href[1]{\endgroup#1\@@endlink}%
\providecommand \@sanitize@url [0]{\catcode `\\12\catcode `\$12\catcode
  `\&12\catcode `\#12\catcode `\^12\catcode `\_12\catcode `\%12\relax}%
\providecommand \@@startlink[1]{}%
\providecommand \@@endlink[0]{}%
\providecommand \url  [0]{\begingroup\@sanitize@url \@url }%
\providecommand \@url [1]{\endgroup\@href {#1}{\urlprefix }}%
\providecommand \urlprefix  [0]{URL }%
\providecommand \Eprint [0]{\href }%
\providecommand \doibase [0]{https://doi.org/}%
\providecommand \selectlanguage [0]{\@gobble}%
\providecommand \bibinfo  [0]{\@secondoftwo}%
\providecommand \bibfield  [0]{\@secondoftwo}%
\providecommand \translation [1]{[#1]}%
\providecommand \BibitemOpen [0]{}%
\providecommand \bibitemStop [0]{}%
\providecommand \bibitemNoStop [0]{.\EOS\space}%
\providecommand \EOS [0]{\spacefactor3000\relax}%
\providecommand \BibitemShut  [1]{\csname bibitem#1\endcsname}%
\let\auto@bib@innerbib\@empty
\bibitem [{\citenamefont {Poynting}(1884)}]{ref1}%
  \BibitemOpen
  \bibfield  {author} {\bibinfo {author} {\bibfnamefont {J.~H.}\ \bibnamefont
  {Poynting}},\ }\bibfield  {title} {\bibinfo {title} {On the transfer of
  energy in the electromagnetic field},\ }\href
  {https://doi.org/10.1098/rstl.1884.0016} {\bibfield  {journal} {\bibinfo
  {journal} {Philos. Trans. R. Soc. Lond.}\ }\textbf {\bibinfo {volume}
  {175}},\ \bibinfo {pages} {343–361} (\bibinfo {year} {1884})}\BibitemShut
  {NoStop}%
\bibitem [{\citenamefont {Lipkin}(1964)}]{ref2}%
  \BibitemOpen
  \bibfield  {author} {\bibinfo {author} {\bibfnamefont {D.~M.}\ \bibnamefont
  {Lipkin}},\ }\bibfield  {title} {\bibinfo {title} {Existence of a new
  conservation law in electromagnetic theory},\ }\href
  {https://doi.org/10.1063/1.1704165} {\bibfield  {journal} {\bibinfo
  {journal} {J. Math. Phys.}\ }\textbf {\bibinfo {volume} {5}},\ \bibinfo
  {pages} {696–700} (\bibinfo {year} {1964})}\BibitemShut {NoStop}%
\bibitem [{\citenamefont {Tang}\ and\ \citenamefont {Cohen}(2010)}]{ref3}%
  \BibitemOpen
  \bibfield  {author} {\bibinfo {author} {\bibfnamefont {Y.}~\bibnamefont
  {Tang}}\ and\ \bibinfo {author} {\bibfnamefont {A.~E.}\ \bibnamefont
  {Cohen}},\ }\bibfield  {title} {\bibinfo {title} {Optical chirality and its
  interaction with matter},\ }\href
  {https://doi.org/10.1103/PhysRevLett.104.163901} {\bibfield  {journal}
  {\bibinfo  {journal} {Phys. Rev. Lett.}\ }\textbf {\bibinfo {volume} {104}},\
  \bibinfo {pages} {163901} (\bibinfo {year} {2010})}\BibitemShut {NoStop}%
\bibitem [{\citenamefont {Darwin}(1932)}]{ref4}%
  \BibitemOpen
  \bibfield  {author} {\bibinfo {author} {\bibfnamefont {C.~G.}\ \bibnamefont
  {Darwin}},\ }\bibfield  {title} {\bibinfo {title} {Notes on the theory of
  radiation},\ }\href {https://doi.org/10.1098/rspa.1932.0065} {\bibfield
  {journal} {\bibinfo  {journal} {Proc. R. Soc. London, Ser. A}\ }\textbf
  {\bibinfo {volume} {136}},\ \bibinfo {pages} {36–52} (\bibinfo {year}
  {1932})}\BibitemShut {NoStop}%
\bibitem [{\citenamefont {Landau}\ and\ \citenamefont {Lifshitz}(1951)}]{ref5}%
  \BibitemOpen
  \bibfield  {author} {\bibinfo {author} {\bibfnamefont {L.~D.}\ \bibnamefont
  {Landau}}\ and\ \bibinfo {author} {\bibfnamefont {E.~M.}\ \bibnamefont
  {Lifshitz}},\ }\href@noop {} {\emph {\bibinfo {title} {The Classical Theory
  of Fields}}}\ (\bibinfo  {publisher} {Pergamon Press},\ \bibinfo {address}
  {Oxford},\ \bibinfo {year} {1951})\BibitemShut {NoStop}%
\bibitem [{\citenamefont {Jackson}(1962)}]{ref6}%
  \BibitemOpen
  \bibfield  {author} {\bibinfo {author} {\bibfnamefont {J.~D.}\ \bibnamefont
  {Jackson}},\ }\href@noop {} {\emph {\bibinfo {title} {Classical
  Electrodynamics}}}\ (\bibinfo  {publisher} {John Wiley and Sons, Inc.},\
  \bibinfo {address} {New York},\ \bibinfo {year} {1962})\BibitemShut {NoStop}%
\bibitem [{\citenamefont {Jones}\ \emph {et~al.}(2015)\citenamefont {Jones},
  \citenamefont {Maragò},\ and\ \citenamefont {Volpe}}]{ref7}%
  \BibitemOpen
  \bibfield  {author} {\bibinfo {author} {\bibfnamefont {P.~H.}\ \bibnamefont
  {Jones}}, \bibinfo {author} {\bibfnamefont {O.~M.}\ \bibnamefont {Maragò}},\
  and\ \bibinfo {author} {\bibfnamefont {G.}~\bibnamefont {Volpe}},\
  }\href@noop {} {\emph {\bibinfo {title} {Optical Tweezers: Principles and
  Applications}}}\ (\bibinfo  {publisher} {Cambridge University Press},\
  \bibinfo {address} {Cambridge},\ \bibinfo {year} {2015})\BibitemShut
  {NoStop}%
\bibitem [{\citenamefont {Poynting}(1909)}]{ref8}%
  \BibitemOpen
  \bibfield  {author} {\bibinfo {author} {\bibfnamefont {J.~H.}\ \bibnamefont
  {Poynting}},\ }\bibfield  {title} {\bibinfo {title} {The wave motion of a
  revolving shaft, and a suggestion as to the angular momentum in a beam of
  circularly polarised light},\ }\href {https://doi.org/10.1098/rspa.1909.0060}
  {\bibfield  {journal} {\bibinfo  {journal} {Proc. R. Soc. London, Ser. A}\
  }\textbf {\bibinfo {volume} {82}},\ \bibinfo {pages} {560–567} (\bibinfo
  {year} {1909})}\BibitemShut {NoStop}%
\bibitem [{\citenamefont {Beth}(1936)}]{ref9}%
  \BibitemOpen
  \bibfield  {author} {\bibinfo {author} {\bibfnamefont {R.~A.}\ \bibnamefont
  {Beth}},\ }\bibfield  {title} {\bibinfo {title} {Mechanical detection and
  measurement of the angular momentum of light},\ }\href
  {https://doi.org/10.1103/PhysRev.50.115} {\bibfield  {journal} {\bibinfo
  {journal} {Phys. Rev.}\ }\textbf {\bibinfo {volume} {50}},\ \bibinfo {pages}
  {115–125} (\bibinfo {year} {1936})}\BibitemShut {NoStop}%
\bibitem [{\citenamefont {Marston}\ and\ \citenamefont
  {Crichton}(1984)}]{ref10}%
  \BibitemOpen
  \bibfield  {author} {\bibinfo {author} {\bibfnamefont {P.~L.}\ \bibnamefont
  {Marston}}\ and\ \bibinfo {author} {\bibfnamefont {J.~H.}\ \bibnamefont
  {Crichton}},\ }\bibfield  {title} {\bibinfo {title} {Radiation torque on a
  sphere caused by a circularly-polarized electromagnetic wave},\ }\href
  {https://doi.org/10.1103/PhysRevA.30.2508} {\bibfield  {journal} {\bibinfo
  {journal} {Phys. Rev. A}\ }\textbf {\bibinfo {volume} {30}},\ \bibinfo
  {pages} {2508–2516} (\bibinfo {year} {1984})}\BibitemShut {NoStop}%
\bibitem [{\citenamefont {Santamato}\ \emph {et~al.}(1986)\citenamefont
  {Santamato}, \citenamefont {Daino}, \citenamefont {Romagnoli}, \citenamefont
  {Settembre},\ and\ \citenamefont {Shen}}]{ref11}%
  \BibitemOpen
  \bibfield  {author} {\bibinfo {author} {\bibfnamefont {E.}~\bibnamefont
  {Santamato}}, \bibinfo {author} {\bibfnamefont {B.}~\bibnamefont {Daino}},
  \bibinfo {author} {\bibfnamefont {M.}~\bibnamefont {Romagnoli}}, \bibinfo
  {author} {\bibfnamefont {M.}~\bibnamefont {Settembre}},\ and\ \bibinfo
  {author} {\bibfnamefont {Y.~R.}\ \bibnamefont {Shen}},\ }\bibfield  {title}
  {\bibinfo {title} {Radiation torque on a sphere caused by a
  circularly-polarized electromagnetic wave},\ }\href
  {https://doi.org/10.1103/PhysRevLett.57.2423} {\bibfield  {journal} {\bibinfo
   {journal} {Phys. Rev. Lett.}\ }\textbf {\bibinfo {volume} {57}},\ \bibinfo
  {pages} {2423–2426} (\bibinfo {year} {1986})}\BibitemShut {NoStop}%
\bibitem [{\citenamefont {Sugiura}\ \emph {et~al.}(1990)\citenamefont
  {Sugiura}, \citenamefont {Kawata},\ and\ \citenamefont {Minami}}]{ref12}%
  \BibitemOpen
  \bibfield  {author} {\bibinfo {author} {\bibfnamefont {T.}~\bibnamefont
  {Sugiura}}, \bibinfo {author} {\bibfnamefont {S.}~\bibnamefont {Kawata}},\
  and\ \bibinfo {author} {\bibfnamefont {S.}~\bibnamefont {Minami}},\
  }\bibfield  {title} {\bibinfo {title} {Optical rotation of small particles by
  a circularly-polarized laser beam in an optical microscope},\ }\href
  {https://doi.org/10.5111/bunkou.39.342} {\bibfield  {journal} {\bibinfo
  {journal} {J. Spectrosc. Soc. Jpn.}\ }\textbf {\bibinfo {volume} {39}},\
  \bibinfo {pages} {342–346} (\bibinfo {year} {1990})}\BibitemShut {NoStop}%
\bibitem [{\citenamefont {Friese}\ \emph {et~al.}(1998)\citenamefont {Friese},
  \citenamefont {Nieminen}, \citenamefont {Heckenberg},\ and\ \citenamefont
  {Rubinsztein-Dunlop}}]{ref13}%
  \BibitemOpen
  \bibfield  {author} {\bibinfo {author} {\bibfnamefont {M.~E.~J.}\
  \bibnamefont {Friese}}, \bibinfo {author} {\bibfnamefont {T.~A.}\
  \bibnamefont {Nieminen}}, \bibinfo {author} {\bibfnamefont {N.~R.}\
  \bibnamefont {Heckenberg}},\ and\ \bibinfo {author} {\bibfnamefont
  {H.}~\bibnamefont {Rubinsztein-Dunlop}},\ }\bibfield  {title} {\bibinfo
  {title} {Optical rotation of small particles by a circularly-polarized laser
  beam in an optical microscope},\ }\href
  {https://doi.org/10.1364/OL.23.000001} {\bibfield  {journal} {\bibinfo
  {journal} {Opt. Lett.}\ }\textbf {\bibinfo {volume} {23}},\ \bibinfo {pages}
  {1–3} (\bibinfo {year} {1998})}\BibitemShut {NoStop}%
\bibitem [{\citenamefont {He}\ \emph {et~al.}(1995{\natexlab{a}})\citenamefont
  {He}, \citenamefont {Heckenberg},\ and\ \citenamefont
  {Rubinsztein-Dunlop}}]{ref14}%
  \BibitemOpen
  \bibfield  {author} {\bibinfo {author} {\bibfnamefont {H.}~\bibnamefont
  {He}}, \bibinfo {author} {\bibfnamefont {N.~R.}\ \bibnamefont {Heckenberg}},\
  and\ \bibinfo {author} {\bibfnamefont {H.}~\bibnamefont
  {Rubinsztein-Dunlop}},\ }\bibfield  {title} {\bibinfo {title} {Optical
  particle trapping with higher-order doughnut beams produced using high
  efficiency computer generated holograms},\ }\href
  {https://doi.org/10.1080/09500349514550171} {\bibfield  {journal} {\bibinfo
  {journal} {J. Mod. Opt.}\ }\textbf {\bibinfo {volume} {42}},\ \bibinfo
  {pages} {217–223} (\bibinfo {year} {1995}{\natexlab{a}})}\BibitemShut
  {NoStop}%
\bibitem [{\citenamefont {He}\ \emph {et~al.}(1995{\natexlab{b}})\citenamefont
  {He}, \citenamefont {Friese}, \citenamefont {Heckenberg},\ and\ \citenamefont
  {Rubinsztein-Dunlop}}]{ref15}%
  \BibitemOpen
  \bibfield  {author} {\bibinfo {author} {\bibfnamefont {H.}~\bibnamefont
  {He}}, \bibinfo {author} {\bibfnamefont {M.~E.~J.}\ \bibnamefont {Friese}},
  \bibinfo {author} {\bibfnamefont {N.~R.}\ \bibnamefont {Heckenberg}},\ and\
  \bibinfo {author} {\bibfnamefont {H.}~\bibnamefont {Rubinsztein-Dunlop}},\
  }\bibfield  {title} {\bibinfo {title} {Direct observation of transfer of
  angular momentum to absorptive particles from a laser beam with a phase
  singularity},\ }\href {https://doi.org/10.1103/PhysRevLett.75.826} {\bibfield
   {journal} {\bibinfo  {journal} {Phys. Rev. Lett.}\ }\textbf {\bibinfo
  {volume} {75}},\ \bibinfo {pages} {826–829} (\bibinfo {year}
  {1995}{\natexlab{b}})}\BibitemShut {NoStop}%
\bibitem [{\citenamefont {Simpson}\ \emph {et~al.}(1996)\citenamefont
  {Simpson}, \citenamefont {Allen},\ and\ \citenamefont {Padgett}}]{ref16}%
  \BibitemOpen
  \bibfield  {author} {\bibinfo {author} {\bibfnamefont {N.~B.}\ \bibnamefont
  {Simpson}}, \bibinfo {author} {\bibfnamefont {L.}~\bibnamefont {Allen}},\
  and\ \bibinfo {author} {\bibfnamefont {M.~J.}\ \bibnamefont {Padgett}},\
  }\bibfield  {title} {\bibinfo {title} {Optical tweezers and optical spanners
  with laguerre–gaussian modes},\ }\href
  {https://doi.org/10.1080/09500349608230675} {\bibfield  {journal} {\bibinfo
  {journal} {J. Mod. Opt.}\ }\textbf {\bibinfo {volume} {43}},\ \bibinfo
  {pages} {2485–2491} (\bibinfo {year} {1996})}\BibitemShut {NoStop}%
\bibitem [{\citenamefont {Simpson}\ \emph {et~al.}(1997)\citenamefont
  {Simpson}, \citenamefont {Dholakia}, \citenamefont {Allen},\ and\
  \citenamefont {Padgett}}]{ref17}%
  \BibitemOpen
  \bibfield  {author} {\bibinfo {author} {\bibfnamefont {N.~B.}\ \bibnamefont
  {Simpson}}, \bibinfo {author} {\bibfnamefont {K.}~\bibnamefont {Dholakia}},
  \bibinfo {author} {\bibfnamefont {L.}~\bibnamefont {Allen}},\ and\ \bibinfo
  {author} {\bibfnamefont {M.~J.}\ \bibnamefont {Padgett}},\ }\bibfield
  {title} {\bibinfo {title} {Mechanical equivalence of spin and orbital angular
  momentum of light: an optical spanner},\ }\href
  {https://doi.org/10.1364/OL.22.000052} {\bibfield  {journal} {\bibinfo
  {journal} {Opt. Lett.}\ }\textbf {\bibinfo {volume} {22}},\ \bibinfo {pages}
  {52–54} (\bibinfo {year} {1997})}\BibitemShut {NoStop}%
\bibitem [{\citenamefont {Garcés-Chávez}\ \emph {et~al.}(2003)\citenamefont
  {Garcés-Chávez}, \citenamefont {McGloin}, \citenamefont {Padgett},
  \citenamefont {Dultz}, \citenamefont {Schmitzer},\ and\ \citenamefont
  {Dholakia}}]{ref18}%
  \BibitemOpen
  \bibfield  {author} {\bibinfo {author} {\bibfnamefont {V.}~\bibnamefont
  {Garcés-Chávez}}, \bibinfo {author} {\bibfnamefont {D.}~\bibnamefont
  {McGloin}}, \bibinfo {author} {\bibfnamefont {M.}~\bibnamefont {Padgett}},
  \bibinfo {author} {\bibfnamefont {W.}~\bibnamefont {Dultz}}, \bibinfo
  {author} {\bibfnamefont {H.}~\bibnamefont {Schmitzer}},\ and\ \bibinfo
  {author} {\bibfnamefont {K.}~\bibnamefont {Dholakia}},\ }\bibfield  {title}
  {\bibinfo {title} {Observation of the transfer of the local angular momentum
  density of a multiringed light beam to an optically trapped particle},\
  }\href {https://doi.org/10.1103/PhysRevLett.91.093602} {\bibfield  {journal}
  {\bibinfo  {journal} {Phys. Rev. Lett.}\ }\textbf {\bibinfo {volume} {91}},\
  \bibinfo {pages} {093602} (\bibinfo {year} {2003})}\BibitemShut {NoStop}%
\bibitem [{\citenamefont {Zhao}\ \emph {et~al.}(2009)\citenamefont {Zhao},
  \citenamefont {Shapiro}, \citenamefont {Mcgloin}, \citenamefont {Chiu},\ and\
  \citenamefont {Marchesini}}]{ref19}%
  \BibitemOpen
  \bibfield  {author} {\bibinfo {author} {\bibfnamefont {Y.}~\bibnamefont
  {Zhao}}, \bibinfo {author} {\bibfnamefont {D.}~\bibnamefont {Shapiro}},
  \bibinfo {author} {\bibfnamefont {D.}~\bibnamefont {Mcgloin}}, \bibinfo
  {author} {\bibfnamefont {D.}~\bibnamefont {Chiu}},\ and\ \bibinfo {author}
  {\bibfnamefont {S.}~\bibnamefont {Marchesini}},\ }\bibfield  {title}
  {\bibinfo {title} {Direct observation of the transfer of orbital angular
  momentum to metal particles from a focused circularly polarized gaussian
  beam},\ }\href {https://doi.org/10.1364/OE.17.023316} {\bibfield  {journal}
  {\bibinfo  {journal} {Opt. Express}\ }\textbf {\bibinfo {volume} {17}},\
  \bibinfo {pages} {23316} (\bibinfo {year} {2009})}\BibitemShut {NoStop}%
\bibitem [{\citenamefont {Fujiwara}\ \emph {et~al.}(2021)\citenamefont
  {Fujiwara}, \citenamefont {Sudo}, \citenamefont {Sunaba}, \citenamefont
  {Pin}, \citenamefont {Ishida},\ and\ \citenamefont {Sasaki}}]{ref20}%
  \BibitemOpen
  \bibfield  {author} {\bibinfo {author} {\bibfnamefont {H.}~\bibnamefont
  {Fujiwara}}, \bibinfo {author} {\bibfnamefont {K.}~\bibnamefont {Sudo}},
  \bibinfo {author} {\bibfnamefont {Y.}~\bibnamefont {Sunaba}}, \bibinfo
  {author} {\bibfnamefont {C.}~\bibnamefont {Pin}}, \bibinfo {author}
  {\bibfnamefont {S.}~\bibnamefont {Ishida}},\ and\ \bibinfo {author}
  {\bibfnamefont {K.}~\bibnamefont {Sasaki}},\ }\bibfield  {title} {\bibinfo
  {title} {Direct observation of the transfer of orbital angular momentum to
  metal particles from a focused circularly polarized gaussian beam},\ }\href
  {https://doi.org/10.1021/acs.nanolett.1c02083} {\bibfield  {journal}
  {\bibinfo  {journal} {Nano Lett.}\ }\textbf {\bibinfo {volume} {21}},\
  \bibinfo {pages} {6268–6273} (\bibinfo {year} {2021})}\BibitemShut
  {NoStop}%
\bibitem [{\citenamefont {Rivera}\ \emph {et~al.}(2016)\citenamefont {Rivera},
  \citenamefont {Seyler}, \citenamefont {Yu}, \citenamefont {Schaibley},
  \citenamefont {Yan}, \citenamefont {Mandrus}, \citenamefont {Yao},\ and\
  \citenamefont {Xu}}]{ref21}%
  \BibitemOpen
  \bibfield  {author} {\bibinfo {author} {\bibfnamefont {P.}~\bibnamefont
  {Rivera}}, \bibinfo {author} {\bibfnamefont {K.}~\bibnamefont {Seyler}},
  \bibinfo {author} {\bibfnamefont {H.}~\bibnamefont {Yu}}, \bibinfo {author}
  {\bibfnamefont {J.}~\bibnamefont {Schaibley}}, \bibinfo {author}
  {\bibfnamefont {J.}~\bibnamefont {Yan}}, \bibinfo {author} {\bibfnamefont
  {D.}~\bibnamefont {Mandrus}}, \bibinfo {author} {\bibfnamefont
  {W.}~\bibnamefont {Yao}},\ and\ \bibinfo {author} {\bibfnamefont
  {X.}~\bibnamefont {Xu}},\ }\bibfield  {title} {\bibinfo {title}
  {Valley-polarized exciton dynamics in a 2d semiconductor heterostructure},\
  }\href {https://doi.org/10.1126/science.aac7820} {\bibfield  {journal}
  {\bibinfo  {journal} {Science}\ }\textbf {\bibinfo {volume} {351}},\ \bibinfo
  {pages} {688–691} (\bibinfo {year} {2016})}\BibitemShut {NoStop}%
\bibitem [{\citenamefont {Schmiegelow}\ \emph {et~al.}(2016)\citenamefont
  {Schmiegelow}, \citenamefont {Schulz}, \citenamefont {Kaufmann},
  \citenamefont {Ruster}, \citenamefont {Poschinger},\ and\ \citenamefont
  {Schmidt-Kaler}}]{ref22}%
  \BibitemOpen
  \bibfield  {author} {\bibinfo {author} {\bibfnamefont {C.~T.}\ \bibnamefont
  {Schmiegelow}}, \bibinfo {author} {\bibfnamefont {J.}~\bibnamefont {Schulz}},
  \bibinfo {author} {\bibfnamefont {H.}~\bibnamefont {Kaufmann}}, \bibinfo
  {author} {\bibfnamefont {T.}~\bibnamefont {Ruster}}, \bibinfo {author}
  {\bibfnamefont {U.~G.}\ \bibnamefont {Poschinger}},\ and\ \bibinfo {author}
  {\bibfnamefont {F.}~\bibnamefont {Schmidt-Kaler}},\ }\bibfield  {title}
  {\bibinfo {title} {Transfer of optical orbital angular momentum to a bound
  electron},\ }\href {https://doi.org/10.1038/ncomms12998} {\bibfield
  {journal} {\bibinfo  {journal} {Nat. Commun.}\ }\textbf {\bibinfo {volume}
  {7}},\ \bibinfo {pages} {12998} (\bibinfo {year} {2016})}\BibitemShut
  {NoStop}%
\bibitem [{\citenamefont {Arikawa}\ \emph {et~al.}(2020)\citenamefont
  {Arikawa}, \citenamefont {Hiraoka}, \citenamefont {Morimoto}, \citenamefont
  {Blanchard}, \citenamefont {Tani}, \citenamefont {Tanaka}, \citenamefont
  {Sakai}, \citenamefont {Kitajima}, \citenamefont {Sasaki},\ and\
  \citenamefont {Tanaka}}]{ref23}%
  \BibitemOpen
  \bibfield  {author} {\bibinfo {author} {\bibfnamefont {T.}~\bibnamefont
  {Arikawa}}, \bibinfo {author} {\bibfnamefont {T.}~\bibnamefont {Hiraoka}},
  \bibinfo {author} {\bibfnamefont {S.}~\bibnamefont {Morimoto}}, \bibinfo
  {author} {\bibfnamefont {F.}~\bibnamefont {Blanchard}}, \bibinfo {author}
  {\bibfnamefont {S.}~\bibnamefont {Tani}}, \bibinfo {author} {\bibfnamefont
  {T.}~\bibnamefont {Tanaka}}, \bibinfo {author} {\bibfnamefont
  {K.}~\bibnamefont {Sakai}}, \bibinfo {author} {\bibfnamefont
  {H.}~\bibnamefont {Kitajima}}, \bibinfo {author} {\bibfnamefont
  {K.}~\bibnamefont {Sasaki}},\ and\ \bibinfo {author} {\bibfnamefont
  {K.}~\bibnamefont {Tanaka}},\ }\bibfield  {title} {\bibinfo {title} {Transfer
  of orbital angular momentum of light to plasmonic excitations in
  metamaterials},\ }\href {https://doi.org/10.1126/sciadv.aay1977} {\bibfield
  {journal} {\bibinfo  {journal} {Sci. Adv.}\ }\textbf {\bibinfo {volume}
  {6}},\ \bibinfo {pages} {eaay1977} (\bibinfo {year} {2020})}\BibitemShut
  {NoStop}%
\bibitem [{\citenamefont {Papakostas}\ \emph {et~al.}(2003)\citenamefont
  {Papakostas}, \citenamefont {Potts}, \citenamefont {Bagnall}, \citenamefont
  {Prosvirnin}, \citenamefont {Coles},\ and\ \citenamefont {Zheludev}}]{ref24}%
  \BibitemOpen
  \bibfield  {author} {\bibinfo {author} {\bibfnamefont {A.}~\bibnamefont
  {Papakostas}}, \bibinfo {author} {\bibfnamefont {A.}~\bibnamefont {Potts}},
  \bibinfo {author} {\bibfnamefont {D.~M.}\ \bibnamefont {Bagnall}}, \bibinfo
  {author} {\bibfnamefont {S.~L.}\ \bibnamefont {Prosvirnin}}, \bibinfo
  {author} {\bibfnamefont {H.~J.}\ \bibnamefont {Coles}},\ and\ \bibinfo
  {author} {\bibfnamefont {N.~I.}\ \bibnamefont {Zheludev}},\ }\bibfield
  {title} {\bibinfo {title} {Transfer of orbital angular momentum of light to
  plasmonic excitations in metamaterials},\ }\href
  {https://doi.org/10.1103/PhysRevLett.90.107404} {\bibfield  {journal}
  {\bibinfo  {journal} {Phys. Rev. Lett.}\ }\textbf {\bibinfo {volume} {90}},\
  \bibinfo {pages} {107404} (\bibinfo {year} {2003})}\BibitemShut {NoStop}%
\bibitem [{\citenamefont {Vallius}\ \emph {et~al.}(2003)\citenamefont
  {Vallius}, \citenamefont {Jefimovs}, \citenamefont {Turunen}, \citenamefont
  {Vahimaa},\ and\ \citenamefont {Svirko}}]{ref25}%
  \BibitemOpen
  \bibfield  {author} {\bibinfo {author} {\bibfnamefont {T.}~\bibnamefont
  {Vallius}}, \bibinfo {author} {\bibfnamefont {K.}~\bibnamefont {Jefimovs}},
  \bibinfo {author} {\bibfnamefont {J.}~\bibnamefont {Turunen}}, \bibinfo
  {author} {\bibfnamefont {P.}~\bibnamefont {Vahimaa}},\ and\ \bibinfo {author}
  {\bibfnamefont {Y.}~\bibnamefont {Svirko}},\ }\bibfield  {title} {\bibinfo
  {title} {Optical activity in subwavelength-period arrays of chiral metallic
  particles},\ }\href {https://doi.org/10.1063/1.1592015} {\bibfield  {journal}
  {\bibinfo  {journal} {Appl. Phys. Lett.}\ }\textbf {\bibinfo {volume} {83}},\
  \bibinfo {pages} {234–236} (\bibinfo {year} {2003})}\BibitemShut {NoStop}%
\bibitem [{\citenamefont {Zhang}\ \emph {et~al.}(2011)\citenamefont {Zhang},
  \citenamefont {Wei}, \citenamefont {Bao}, \citenamefont {Håkanson},
  \citenamefont {Halas}, \citenamefont {Nordlander},\ and\ \citenamefont
  {Xu}}]{ref26}%
  \BibitemOpen
  \bibfield  {author} {\bibinfo {author} {\bibfnamefont {S.}~\bibnamefont
  {Zhang}}, \bibinfo {author} {\bibfnamefont {H.}~\bibnamefont {Wei}}, \bibinfo
  {author} {\bibfnamefont {K.}~\bibnamefont {Bao}}, \bibinfo {author}
  {\bibfnamefont {U.}~\bibnamefont {Håkanson}}, \bibinfo {author}
  {\bibfnamefont {N.~J.}\ \bibnamefont {Halas}}, \bibinfo {author}
  {\bibfnamefont {P.}~\bibnamefont {Nordlander}},\ and\ \bibinfo {author}
  {\bibfnamefont {H.}~\bibnamefont {Xu}},\ }\bibfield  {title} {\bibinfo
  {title} {Chiral surface plasmon polaritons on metallic nanowires},\ }\href
  {https://doi.org/10.1103/PhysRevLett.107.096801} {\bibfield  {journal}
  {\bibinfo  {journal} {Phys. Rev. Lett.}\ }\textbf {\bibinfo {volume} {107}},\
  \bibinfo {pages} {096801} (\bibinfo {year} {2011})}\BibitemShut {NoStop}%
\bibitem [{\citenamefont {Zhang}\ and\ \citenamefont {Niu}(2014)}]{ref27}%
  \BibitemOpen
  \bibfield  {author} {\bibinfo {author} {\bibfnamefont {L.}~\bibnamefont
  {Zhang}}\ and\ \bibinfo {author} {\bibfnamefont {Q.}~\bibnamefont {Niu}},\
  }\bibfield  {title} {\bibinfo {title} {Angular momentum of phonons and the
  einstein–de haas effect},\ }\href
  {https://doi.org/10.1103/PhysRevLett.112.085503} {\bibfield  {journal}
  {\bibinfo  {journal} {Phys. Rev. Lett.}\ }\textbf {\bibinfo {volume} {112}},\
  \bibinfo {pages} {085503} (\bibinfo {year} {2014})}\BibitemShut {NoStop}%
\bibitem [{\citenamefont {Chen}\ \emph {et~al.}(2019)\citenamefont {Chen},
  \citenamefont {Zhang}, \citenamefont {Niu},\ and\ \citenamefont
  {Zhang}}]{ref28}%
  \BibitemOpen
  \bibfield  {author} {\bibinfo {author} {\bibfnamefont {H.}~\bibnamefont
  {Chen}}, \bibinfo {author} {\bibfnamefont {W.}~\bibnamefont {Zhang}},
  \bibinfo {author} {\bibfnamefont {Q.}~\bibnamefont {Niu}},\ and\ \bibinfo
  {author} {\bibfnamefont {L.}~\bibnamefont {Zhang}},\ }\bibfield  {title}
  {\bibinfo {title} {Chiral phonons in two-dimensional materials},\ }\href
  {https://doi.org/10.1088/2053-1583/aaf292} {\bibfield  {journal} {\bibinfo
  {journal} {2D Mater.}\ }\textbf {\bibinfo {volume} {6}},\ \bibinfo {pages}
  {012002} (\bibinfo {year} {2019})}\BibitemShut {NoStop}%
\bibitem [{\citenamefont {Ishito}\ \emph {et~al.}(2023)\citenamefont {Ishito},
  \citenamefont {Mao}, \citenamefont {Kousaka}, \citenamefont {Togawa},
  \citenamefont {Iwasaki}, \citenamefont {Zhang}, \citenamefont {Murakami},
  \citenamefont {Kishine},\ and\ \citenamefont {Satoh}}]{ref29}%
  \BibitemOpen
  \bibfield  {author} {\bibinfo {author} {\bibfnamefont {K.}~\bibnamefont
  {Ishito}}, \bibinfo {author} {\bibfnamefont {H.}~\bibnamefont {Mao}},
  \bibinfo {author} {\bibfnamefont {Y.}~\bibnamefont {Kousaka}}, \bibinfo
  {author} {\bibfnamefont {Y.}~\bibnamefont {Togawa}}, \bibinfo {author}
  {\bibfnamefont {S.}~\bibnamefont {Iwasaki}}, \bibinfo {author} {\bibfnamefont
  {T.}~\bibnamefont {Zhang}}, \bibinfo {author} {\bibfnamefont
  {S.}~\bibnamefont {Murakami}}, \bibinfo {author} {\bibfnamefont
  {J.}~\bibnamefont {Kishine}},\ and\ \bibinfo {author} {\bibfnamefont
  {T.}~\bibnamefont {Satoh}},\ }\bibfield  {title} {\bibinfo {title} {Truly
  chiral phonons in $\alpha$-hgs},\ }\href
  {https://doi.org/10.1038/s41567-022-01790-x} {\bibfield  {journal} {\bibinfo
  {journal} {Nat. Phys.}\ }\textbf {\bibinfo {volume} {19}},\ \bibinfo {pages}
  {35–39} (\bibinfo {year} {2023})}\BibitemShut {NoStop}%
\bibitem [{\citenamefont {Ueda}\ \emph {et~al.}(2023)\citenamefont {Ueda},
  \citenamefont {García-Fernández}, \citenamefont {Agrestini}, \citenamefont
  {Romao}, \citenamefont {van~den Brink}, \citenamefont {Spaldin},
  \citenamefont {Zhou},\ and\ \citenamefont {Staub}}]{ref30}%
  \BibitemOpen
  \bibfield  {author} {\bibinfo {author} {\bibfnamefont {H.}~\bibnamefont
  {Ueda}}, \bibinfo {author} {\bibfnamefont {M.}~\bibnamefont
  {García-Fernández}}, \bibinfo {author} {\bibfnamefont {S.}~\bibnamefont
  {Agrestini}}, \bibinfo {author} {\bibfnamefont {C.~P.}\ \bibnamefont
  {Romao}}, \bibinfo {author} {\bibfnamefont {J.}~\bibnamefont {van~den
  Brink}}, \bibinfo {author} {\bibfnamefont {N.~A.}\ \bibnamefont {Spaldin}},
  \bibinfo {author} {\bibfnamefont {K.-J.}\ \bibnamefont {Zhou}},\ and\
  \bibinfo {author} {\bibfnamefont {U.}~\bibnamefont {Staub}},\ }\bibfield
  {title} {\bibinfo {title} {Chiral phonons in quartz probed by x-rays},\
  }\href {https://doi.org/10.1038/s41586-023-06016-5} {\bibfield  {journal}
  {\bibinfo  {journal} {Nature (London)}\ }\textbf {\bibinfo {volume} {19}},\
  \bibinfo {pages} {946–950} (\bibinfo {year} {2023})}\BibitemShut {NoStop}%
\bibitem [{\citenamefont {Oishi}\ \emph {et~al.}(2024)\citenamefont {Oishi},
  \citenamefont {Fujii},\ and\ \citenamefont {Koreeda}}]{ref31}%
  \BibitemOpen
  \bibfield  {author} {\bibinfo {author} {\bibfnamefont {E.}~\bibnamefont
  {Oishi}}, \bibinfo {author} {\bibfnamefont {Y.}~\bibnamefont {Fujii}},\ and\
  \bibinfo {author} {\bibfnamefont {A.}~\bibnamefont {Koreeda}},\ }\bibfield
  {title} {\bibinfo {title} {Selective observation of enantiomeric chiral
  phonons in $\alpha$-quartz},\ }\href
  {https://doi.org/10.1103/PhysRevB.109.104306} {\bibfield  {journal} {\bibinfo
   {journal} {Phys. Rev. B}\ }\textbf {\bibinfo {volume} {109}},\ \bibinfo
  {pages} {104306} (\bibinfo {year} {2024})}\BibitemShut {NoStop}%
\bibitem [{\citenamefont {Rodger}\ and\ \citenamefont {Norden}(1997)}]{ref32}%
  \BibitemOpen
  \bibfield  {author} {\bibinfo {author} {\bibfnamefont {A.}~\bibnamefont
  {Rodger}}\ and\ \bibinfo {author} {\bibfnamefont {B.}~\bibnamefont
  {Norden}},\ }\href@noop {} {\emph {\bibinfo {title} {Circular Dichroism and
  Linear Dichroism: A Textbook on Polarized-Light Spectroscop}}}\ (\bibinfo
  {publisher} {Oxford University Press},\ \bibinfo {address} {Oxford},\
  \bibinfo {year} {1997})\BibitemShut {NoStop}%
\bibitem [{\citenamefont {Berova}\ \emph {et~al.}(2000)\citenamefont {Berova},
  \citenamefont {Nakanishi},\ and\ \citenamefont {Woody}}]{ref33}%
  \BibitemOpen
  \bibfield  {author} {\bibinfo {author} {\bibfnamefont {N.}~\bibnamefont
  {Berova}}, \bibinfo {author} {\bibfnamefont {K.}~\bibnamefont {Nakanishi}},\
  and\ \bibinfo {author} {\bibfnamefont {R.~W.}\ \bibnamefont {Woody}},\
  }\href@noop {} {\emph {\bibinfo {title} {Circular Dichroism: Principles and
  Applications}}}\ (\bibinfo  {publisher} {Wiley-VCH},\ \bibinfo {address} {New
  York},\ \bibinfo {year} {2000})\BibitemShut {NoStop}%
\bibitem [{\citenamefont {Barron}(2004)}]{ref34}%
  \BibitemOpen
  \bibfield  {author} {\bibinfo {author} {\bibfnamefont {L.~D.}\ \bibnamefont
  {Barron}},\ }\href@noop {} {\emph {\bibinfo {title} {Molecular Light
  Scattering and Optical Activity, 2nd ed.}}}\ (\bibinfo  {publisher}
  {Cambridge University Press},\ \bibinfo {address} {Cambridge},\ \bibinfo
  {year} {2004})\BibitemShut {NoStop}%
\bibitem [{\citenamefont {Brullot}\ \emph {et~al.}(2016)\citenamefont
  {Brullot}, \citenamefont {Vanbel}, \citenamefont {Swusten},\ and\
  \citenamefont {Verbiest}}]{ref35}%
  \BibitemOpen
  \bibfield  {author} {\bibinfo {author} {\bibfnamefont {W.}~\bibnamefont
  {Brullot}}, \bibinfo {author} {\bibfnamefont {M.~K.}\ \bibnamefont {Vanbel}},
  \bibinfo {author} {\bibfnamefont {T.}~\bibnamefont {Swusten}},\ and\ \bibinfo
  {author} {\bibfnamefont {T.}~\bibnamefont {Verbiest}},\ }\bibfield  {title}
  {\bibinfo {title} {Resolving enantiomers using the optical angular momentum
  of twisted light},\ }\href {https://doi.org/10.1126/sciadv.1501349}
  {\bibfield  {journal} {\bibinfo  {journal} {Sci. Adv.}\ }\textbf {\bibinfo
  {volume} {2}},\ \bibinfo {pages} {e1501349} (\bibinfo {year}
  {2016})}\BibitemShut {NoStop}%
\bibitem [{\citenamefont {Forbes}\ and\ \citenamefont {Andrews}(2021)}]{ref36}%
  \BibitemOpen
  \bibfield  {author} {\bibinfo {author} {\bibfnamefont {K.~A.}\ \bibnamefont
  {Forbes}}\ and\ \bibinfo {author} {\bibfnamefont {D.~L.}\ \bibnamefont
  {Andrews}},\ }\bibfield  {title} {\bibinfo {title} {Orbital angular momentum
  of twisted light: Chirality and optical activity},\ }\href
  {https://doi.org/10.1088/2515-7647/abdb06} {\bibfield  {journal} {\bibinfo
  {journal} {J. Phys. Photonics}\ }\textbf {\bibinfo {volume} {3}},\ \bibinfo
  {pages} {022007} (\bibinfo {year} {2021})}\BibitemShut {NoStop}%
\bibitem [{\citenamefont {Allen}\ \emph {et~al.}(1992)\citenamefont {Allen},
  \citenamefont {Beijersbergen}, \citenamefont {Spreeuw},\ and\ \citenamefont
  {Woerdman}}]{ref37}%
  \BibitemOpen
  \bibfield  {author} {\bibinfo {author} {\bibfnamefont {L.}~\bibnamefont
  {Allen}}, \bibinfo {author} {\bibfnamefont {M.~W.}\ \bibnamefont
  {Beijersbergen}}, \bibinfo {author} {\bibfnamefont {R.~J.~C.}\ \bibnamefont
  {Spreeuw}},\ and\ \bibinfo {author} {\bibfnamefont {J.~P.}\ \bibnamefont
  {Woerdman}},\ }\bibfield  {title} {\bibinfo {title} {Orbital angular momentum
  of light and the transformation of laguerre-gaussian laser modes},\ }\href
  {https://doi.org/10.1103/PhysRevA.45.8185} {\bibfield  {journal} {\bibinfo
  {journal} {Phys. Rev. A}\ }\textbf {\bibinfo {volume} {45}},\ \bibinfo
  {pages} {8185} (\bibinfo {year} {1992})}\BibitemShut {NoStop}%
\bibitem [{\citenamefont {Marrucci}\ \emph {et~al.}(2006)\citenamefont
  {Marrucci}, \citenamefont {Manzo},\ and\ \citenamefont {Paparo}}]{ref38}%
  \BibitemOpen
  \bibfield  {author} {\bibinfo {author} {\bibfnamefont {L.}~\bibnamefont
  {Marrucci}}, \bibinfo {author} {\bibfnamefont {C.}~\bibnamefont {Manzo}},\
  and\ \bibinfo {author} {\bibfnamefont {D.}~\bibnamefont {Paparo}},\
  }\bibfield  {title} {\bibinfo {title} {Optical spin-to-orbital angular
  momentum conversion in inhomogeneous anisotropic media},\ }\href
  {https://doi.org/10.1103/PhysRevLett.96.163905} {\bibfield  {journal}
  {\bibinfo  {journal} {Phys. Rev. Lett.}\ }\textbf {\bibinfo {volume} {96}},\
  \bibinfo {pages} {163905} (\bibinfo {year} {2006})}\BibitemShut {NoStop}%
\bibitem [{\citenamefont {Zhao}\ \emph {et~al.}(2007)\citenamefont {Zhao},
  \citenamefont {Edgar}, \citenamefont {Jeffries}, \citenamefont {McGloin},\
  and\ \citenamefont {Chiu}}]{ref39}%
  \BibitemOpen
  \bibfield  {author} {\bibinfo {author} {\bibfnamefont {Y.}~\bibnamefont
  {Zhao}}, \bibinfo {author} {\bibfnamefont {J.}~\bibnamefont {Edgar}},
  \bibinfo {author} {\bibfnamefont {G.}~\bibnamefont {Jeffries}}, \bibinfo
  {author} {\bibfnamefont {D.}~\bibnamefont {McGloin}},\ and\ \bibinfo {author}
  {\bibfnamefont {D.}~\bibnamefont {Chiu}},\ }\bibfield  {title} {\bibinfo
  {title} {Spin-to-orbital angular momentum conversion in a strongly focused
  optical beam},\ }\href {https://doi.org/10.1103/PhysRevLett.99.073901}
  {\bibfield  {journal} {\bibinfo  {journal} {Phys. Rev. Lett.}\ }\textbf
  {\bibinfo {volume} {99}},\ \bibinfo {pages} {073901} (\bibinfo {year}
  {2007})}\BibitemShut {NoStop}%
\bibitem [{\citenamefont {Bliokh}\ \emph
  {et~al.}(2010{\natexlab{a}})\citenamefont {Bliokh}, \citenamefont {Alonso},
  \citenamefont {Ostrovskaya},\ and\ \citenamefont {Aiello}}]{ref40}%
  \BibitemOpen
  \bibfield  {author} {\bibinfo {author} {\bibfnamefont {K.~Y.}\ \bibnamefont
  {Bliokh}}, \bibinfo {author} {\bibfnamefont {M.~A.}\ \bibnamefont {Alonso}},
  \bibinfo {author} {\bibfnamefont {E.~A.}\ \bibnamefont {Ostrovskaya}},\ and\
  \bibinfo {author} {\bibfnamefont {A.}~\bibnamefont {Aiello}},\ }\bibfield
  {title} {\bibinfo {title} {Angular momenta and spin-orbit interaction of
  nonparaxial light in free space},\ }\href
  {https://doi.org/10.1103/PhysRevA.82.063825} {\bibfield  {journal} {\bibinfo
  {journal} {Phys. Rev. A}\ }\textbf {\bibinfo {volume} {82}},\ \bibinfo
  {pages} {063825} (\bibinfo {year} {2010}{\natexlab{a}})}\BibitemShut
  {NoStop}%
\bibitem [{\citenamefont {Bliokh}\ \emph
  {et~al.}(2010{\natexlab{b}})\citenamefont {Bliokh}, \citenamefont
  {Ostrovskaya}, \citenamefont {Alonso}, \citenamefont {Rodriguez-Herrera},
  \citenamefont {Lara},\ and\ \citenamefont {Dainty}}]{ref41}%
  \BibitemOpen
  \bibfield  {author} {\bibinfo {author} {\bibfnamefont {K.~Y.}\ \bibnamefont
  {Bliokh}}, \bibinfo {author} {\bibfnamefont {E.~A.}\ \bibnamefont
  {Ostrovskaya}}, \bibinfo {author} {\bibfnamefont {M.~A.}\ \bibnamefont
  {Alonso}}, \bibinfo {author} {\bibfnamefont {O.~G.}\ \bibnamefont
  {Rodriguez-Herrera}}, \bibinfo {author} {\bibfnamefont {D.}~\bibnamefont
  {Lara}},\ and\ \bibinfo {author} {\bibfnamefont {C.}~\bibnamefont {Dainty}},\
  }\bibfield  {title} {\bibinfo {title} {Spin-to-orbital angular momentum
  conversion in focusing, scattering, and imaging systems},\ }\href
  {https://doi.org/10.1364/OE.19.026132} {\bibfield  {journal} {\bibinfo
  {journal} {Opt. Express}\ }\textbf {\bibinfo {volume} {19}},\ \bibinfo
  {pages} {26132} (\bibinfo {year} {2010}{\natexlab{b}})}\BibitemShut {NoStop}%
\bibitem [{\citenamefont {Nechayev}\ \emph {et~al.}(2019)\citenamefont
  {Nechayev}, \citenamefont {Eismann}, \citenamefont {Leuchs},\ and\
  \citenamefont {Banzer}}]{ref42}%
  \BibitemOpen
  \bibfield  {author} {\bibinfo {author} {\bibfnamefont {S.}~\bibnamefont
  {Nechayev}}, \bibinfo {author} {\bibfnamefont {J.~S.}\ \bibnamefont
  {Eismann}}, \bibinfo {author} {\bibfnamefont {G.}~\bibnamefont {Leuchs}},\
  and\ \bibinfo {author} {\bibfnamefont {P.}~\bibnamefont {Banzer}},\
  }\bibfield  {title} {\bibinfo {title} {Orbital-to-spin angular momentum
  conversion employing local helicity},\ }\href
  {https://doi.org/10.1103/PhysRevB.99.075155} {\bibfield  {journal} {\bibinfo
  {journal} {Phys. Rev. B}\ }\textbf {\bibinfo {volume} {99}},\ \bibinfo
  {pages} {075155} (\bibinfo {year} {2019})}\BibitemShut {NoStop}%
\bibitem [{\citenamefont {Yin}\ \emph {et~al.}(2013)\citenamefont {Yin},
  \citenamefont {Ye}, \citenamefont {Rho}, \citenamefont {Wang},\ and\
  \citenamefont {Zhang}}]{refadd1}%
  \BibitemOpen
  \bibfield  {author} {\bibinfo {author} {\bibfnamefont {X.}~\bibnamefont
  {Yin}}, \bibinfo {author} {\bibfnamefont {Z.}~\bibnamefont {Ye}}, \bibinfo
  {author} {\bibfnamefont {J.}~\bibnamefont {Rho}}, \bibinfo {author}
  {\bibfnamefont {Y.}~\bibnamefont {Wang}},\ and\ \bibinfo {author}
  {\bibfnamefont {X.}~\bibnamefont {Zhang}},\ }\bibfield  {title} {\bibinfo
  {title} {Photonic spin hall effect at metasurfaces},\ }\href
  {https://doi.org/10.1126/science.1231758} {\bibfield  {journal} {\bibinfo
  {journal} {Science}\ }\textbf {\bibinfo {volume} {339}},\ \bibinfo {pages}
  {1405–1407} (\bibinfo {year} {2013})}\BibitemShut {NoStop}%
\bibitem [{\citenamefont {Rodríguez-Fortuño}\ \emph
  {et~al.}(2013)\citenamefont {Rodríguez-Fortuño}, \citenamefont {Marino},
  \citenamefont {Ginzburg}, \citenamefont {O’Connor}, \citenamefont
  {Martínez}, \citenamefont {Wurtz},\ and\ \citenamefont {Zayats}}]{refadd2}%
  \BibitemOpen
  \bibfield  {author} {\bibinfo {author} {\bibfnamefont {F.~J.}\ \bibnamefont
  {Rodríguez-Fortuño}}, \bibinfo {author} {\bibfnamefont {G.}~\bibnamefont
  {Marino}}, \bibinfo {author} {\bibfnamefont {P.}~\bibnamefont {Ginzburg}},
  \bibinfo {author} {\bibfnamefont {D.}~\bibnamefont {O’Connor}}, \bibinfo
  {author} {\bibfnamefont {A.}~\bibnamefont {Martínez}}, \bibinfo {author}
  {\bibfnamefont {G.~A.}\ \bibnamefont {Wurtz}},\ and\ \bibinfo {author}
  {\bibfnamefont {A.~V.}\ \bibnamefont {Zayats}},\ }\bibfield  {title}
  {\bibinfo {title} {Near-field interference for the unidirectional excitation
  of electromagnetic guided modes},\ }\href
  {https://doi.org/10.1126/science.1233739} {\bibfield  {journal} {\bibinfo
  {journal} {Science}\ }\textbf {\bibinfo {volume} {340}},\ \bibinfo {pages}
  {328–330} (\bibinfo {year} {2013})}\BibitemShut {NoStop}%
\bibitem [{\citenamefont {Bliokh}\ \emph {et~al.}(2015)\citenamefont {Bliokh},
  \citenamefont {Smirnova},\ and\ \citenamefont {Nori}}]{refadd3}%
  \BibitemOpen
  \bibfield  {author} {\bibinfo {author} {\bibfnamefont {K.~Y.}\ \bibnamefont
  {Bliokh}}, \bibinfo {author} {\bibfnamefont {D.}~\bibnamefont {Smirnova}},\
  and\ \bibinfo {author} {\bibfnamefont {F.}~\bibnamefont {Nori}},\ }\bibfield
  {title} {\bibinfo {title} {Quantum spin hall effect of light},\ }\href
  {https://doi.org/10.1126/science.aaa9519} {\bibfield  {journal} {\bibinfo
  {journal} {Science}\ }\textbf {\bibinfo {volume} {348}},\ \bibinfo {pages}
  {1448–1451} (\bibinfo {year} {2015})}\BibitemShut {NoStop}%
\bibitem [{\citenamefont {Cohen-Tannoudji}\ \emph {et~al.}(1989)\citenamefont
  {Cohen-Tannoudji}, \citenamefont {Dupont-Roc},\ and\ \citenamefont
  {Grynberg}}]{ref43}%
  \BibitemOpen
  \bibfield  {author} {\bibinfo {author} {\bibfnamefont {C.}~\bibnamefont
  {Cohen-Tannoudji}}, \bibinfo {author} {\bibfnamefont {J.}~\bibnamefont
  {Dupont-Roc}},\ and\ \bibinfo {author} {\bibfnamefont {G.}~\bibnamefont
  {Grynberg}},\ }\href@noop {} {\emph {\bibinfo {title} {Photons and Atoms:
  Introduction to Quantum Electrodynamics}}}\ (\bibinfo  {publisher} {Wiley},\
  \bibinfo {address} {New York},\ \bibinfo {year} {1989})\BibitemShut {NoStop}%
\bibitem [{\citenamefont {Crimin}\ \emph {et~al.}(2020)\citenamefont {Crimin},
  \citenamefont {MacKinnon}, \citenamefont {Götte},\ and\ \citenamefont
  {Barnett}}]{ref44}%
  \BibitemOpen
  \bibfield  {author} {\bibinfo {author} {\bibfnamefont {F.}~\bibnamefont
  {Crimin}}, \bibinfo {author} {\bibfnamefont {N.}~\bibnamefont {MacKinnon}},
  \bibinfo {author} {\bibfnamefont {J.}~\bibnamefont {Götte}},\ and\ \bibinfo
  {author} {\bibfnamefont {S.~M.}\ \bibnamefont {Barnett}},\ }\bibfield
  {title} {\bibinfo {title} {Continuous symmetries and conservation laws in
  chiral media},\ }in\ \href@noop {} {\emph {\bibinfo {booktitle} {Proc. SPIE
  11297 Complex Light and Optical Forces XIV}}}\ (\bibinfo {address} {San
  Francisco, USA},\ \bibinfo {year} {2020})\ p.\ \bibinfo {pages}
  {112970J}\BibitemShut {NoStop}%
\bibitem [{\citenamefont {Nieto-Vesperinas}(2015)}]{ref45}%
  \BibitemOpen
  \bibfield  {author} {\bibinfo {author} {\bibfnamefont {M.}~\bibnamefont
  {Nieto-Vesperinas}},\ }\bibfield  {title} {\bibinfo {title} {Optical torque:
  Electromagnetic spin and orbital-angular-momentum conservation laws and their
  significance},\ }\href {https://doi.org/10.1103/PhysRevA.92.043843}
  {\bibfield  {journal} {\bibinfo  {journal} {Phys. Rev. A}\ }\textbf {\bibinfo
  {volume} {92}},\ \bibinfo {pages} {043843} (\bibinfo {year}
  {2015})}\BibitemShut {NoStop}%
\bibitem [{\citenamefont {Nienhuis}(2016)}]{ref46}%
  \BibitemOpen
  \bibfield  {author} {\bibinfo {author} {\bibfnamefont {G.}~\bibnamefont
  {Nienhuis}},\ }\bibfield  {title} {\bibinfo {title} {Conservation laws and
  symmetry transformations of the electromagnetic field with sources},\ }\href
  {https://doi.org/10.1103/PhysRevA.93.023840} {\bibfield  {journal} {\bibinfo
  {journal} {Phys. Rev. A}\ }\textbf {\bibinfo {volume} {93}},\ \bibinfo
  {pages} {023840} (\bibinfo {year} {2016})}\BibitemShut {NoStop}%
\bibitem [{\citenamefont {Bliokh}\ \emph {et~al.}(2014)\citenamefont {Bliokh},
  \citenamefont {Dressel},\ and\ \citenamefont {Nori}}]{ref47}%
  \BibitemOpen
  \bibfield  {author} {\bibinfo {author} {\bibfnamefont {K.~Y.}\ \bibnamefont
  {Bliokh}}, \bibinfo {author} {\bibfnamefont {J.}~\bibnamefont {Dressel}},\
  and\ \bibinfo {author} {\bibfnamefont {F.}~\bibnamefont {Nori}},\ }\bibfield
  {title} {\bibinfo {title} {Conservation of the spin and orbital angular
  momenta in electromagnetism},\ }\href
  {https://doi.org/10.1088/1367-2630/16/9/093037} {\bibfield  {journal}
  {\bibinfo  {journal} {New J. Phys.}\ }\textbf {\bibinfo {volume} {16}},\
  \bibinfo {pages} {093037} (\bibinfo {year} {2014})}\BibitemShut {NoStop}%
\bibitem [{\citenamefont {Feynman}\ \emph {et~al.}(1964)\citenamefont
  {Feynman}, \citenamefont {Leighton},\ and\ \citenamefont {Sands}}]{ref48}%
  \BibitemOpen
  \bibfield  {author} {\bibinfo {author} {\bibfnamefont {R.~P.}\ \bibnamefont
  {Feynman}}, \bibinfo {author} {\bibfnamefont {R.}~\bibnamefont {Leighton}},\
  and\ \bibinfo {author} {\bibfnamefont {M.}~\bibnamefont {Sands}},\
  }\href@noop {} {\emph {\bibinfo {title} {The Feynman Lectures on Physics}}}\
  (\bibinfo  {publisher} {Addison-Wesley Pub. Co.},\ \bibinfo {address}
  {Boston},\ \bibinfo {year} {1964})\BibitemShut {NoStop}%
\bibitem [{\citenamefont {Morgan}(1964)}]{ref51}%
  \BibitemOpen
  \bibfield  {author} {\bibinfo {author} {\bibfnamefont {T.~A.}\ \bibnamefont
  {Morgan}},\ }\bibfield  {title} {\bibinfo {title} {Two classes of new
  conservation laws for the electromagnetic field and for other massless
  fields},\ }\href {https://doi.org/10.1063/1.1931204} {\bibfield  {journal}
  {\bibinfo  {journal} {J. Math. Phys.}\ }\textbf {\bibinfo {volume} {5}},\
  \bibinfo {pages} {1659–1660} (\bibinfo {year} {1964})}\BibitemShut
  {NoStop}%
\bibitem [{\citenamefont {Philbin}(2013)}]{ref52}%
  \BibitemOpen
  \bibfield  {author} {\bibinfo {author} {\bibfnamefont {T.~G.}\ \bibnamefont
  {Philbin}},\ }\bibfield  {title} {\bibinfo {title} {Lipkin’s conservation
  law, noether’s theorem, and the relation to optical helicity},\ }\href
  {https://doi.org/10.1103/PhysRevA.87.043843} {\bibfield  {journal} {\bibinfo
  {journal} {Phys. Rev. A}\ }\textbf {\bibinfo {volume} {87}},\ \bibinfo
  {pages} {043843} (\bibinfo {year} {2013})}\BibitemShut {NoStop}%
\bibitem [{\citenamefont {Proskurin}\ \emph {et~al.}(2017)\citenamefont
  {Proskurin}, \citenamefont {Ovchinnikov}, \citenamefont {Nosov},\ and\
  \citenamefont {Kishine}}]{ref53}%
  \BibitemOpen
  \bibfield  {author} {\bibinfo {author} {\bibfnamefont {I.}~\bibnamefont
  {Proskurin}}, \bibinfo {author} {\bibfnamefont {A.~S.}\ \bibnamefont
  {Ovchinnikov}}, \bibinfo {author} {\bibfnamefont {P.}~\bibnamefont {Nosov}},\
  and\ \bibinfo {author} {\bibfnamefont {J.}~\bibnamefont {Kishine}},\
  }\bibfield  {title} {\bibinfo {title} {Optical chirality in gyrotropic media:
  symmetry approach},\ }\href {https://doi.org/10.1088/1367-2630/aa6acd}
  {\bibfield  {journal} {\bibinfo  {journal} {New J. Phys.}\ }\textbf {\bibinfo
  {volume} {19}},\ \bibinfo {pages} {063021} (\bibinfo {year}
  {2017})}\BibitemShut {NoStop}%
\bibitem [{\citenamefont {Wu}\ \emph {et~al.}(2020)\citenamefont {Wu},
  \citenamefont {Tanaka}, \citenamefont {Fukuhara},\ and\ \citenamefont
  {Shimura}}]{ref49}%
  \BibitemOpen
  \bibfield  {author} {\bibinfo {author} {\bibfnamefont {A.~A.}\ \bibnamefont
  {Wu}}, \bibinfo {author} {\bibfnamefont {Y.~Y.}\ \bibnamefont {Tanaka}},
  \bibinfo {author} {\bibfnamefont {R.}~\bibnamefont {Fukuhara}},\ and\
  \bibinfo {author} {\bibfnamefont {T.}~\bibnamefont {Shimura}},\ }\bibfield
  {title} {\bibinfo {title} {Continuity equation for spin angular momentum in
  relation to optical chirality},\ }\href
  {https://doi.org/10.1103/PhysRevA.102.023531} {\bibfield  {journal} {\bibinfo
   {journal} {Phys. Rev. A}\ }\textbf {\bibinfo {volume} {102}},\ \bibinfo
  {pages} {023531} (\bibinfo {year} {2020})}\BibitemShut {NoStop}%
\bibitem [{\citenamefont {Cho}(2010)}]{refadd5}%
  \BibitemOpen
  \bibfield  {author} {\bibinfo {author} {\bibfnamefont {K.}~\bibnamefont
  {Cho}},\ }\href@noop {} {\emph {\bibinfo {title} {Reconstruction of
  Macroscopic Maxwell Equations: A Single Susceptibility Theory}}}\ (\bibinfo
  {publisher} {Springer Berlin},\ \bibinfo {address} {Heidelberg},\ \bibinfo
  {year} {2010})\BibitemShut {NoStop}%
\bibitem [{\citenamefont {Bohren}\ and\ \citenamefont {Huffman}(2004)}]{ref50}%
  \BibitemOpen
  \bibfield  {author} {\bibinfo {author} {\bibfnamefont {C.~F.}\ \bibnamefont
  {Bohren}}\ and\ \bibinfo {author} {\bibfnamefont {D.~R.}\ \bibnamefont
  {Huffman}},\ }\href@noop {} {\emph {\bibinfo {title} {Absorption and
  Scattering of Light by Small Particles}}}\ (\bibinfo  {publisher}
  {Wiley-VCH},\ \bibinfo {address} {New York},\ \bibinfo {year}
  {2004})\BibitemShut {NoStop}%
\bibitem [{\citenamefont {Tsesses}\ \emph {et~al.}(2018)\citenamefont
  {Tsesses}, \citenamefont {Ostrovsky}, \citenamefont {Cohen}, \citenamefont
  {Gjonaj}, \citenamefont {Lindner},\ and\ \citenamefont {Bartal}}]{refadd4}%
  \BibitemOpen
  \bibfield  {author} {\bibinfo {author} {\bibfnamefont {S.}~\bibnamefont
  {Tsesses}}, \bibinfo {author} {\bibfnamefont {E.}~\bibnamefont {Ostrovsky}},
  \bibinfo {author} {\bibfnamefont {K.}~\bibnamefont {Cohen}}, \bibinfo
  {author} {\bibfnamefont {B.}~\bibnamefont {Gjonaj}}, \bibinfo {author}
  {\bibfnamefont {N.~H.}\ \bibnamefont {Lindner}},\ and\ \bibinfo {author}
  {\bibfnamefont {G.}~\bibnamefont {Bartal}},\ }\bibfield  {title} {\bibinfo
  {title} {Optical skyrmion lattice in evanescent electromagnetic fields},\
  }\href {https://doi.org/10.1126/science.aau0227} {\bibfield  {journal}
  {\bibinfo  {journal} {Science}\ }\textbf {\bibinfo {volume} {361}},\ \bibinfo
  {pages} {993–996} (\bibinfo {year} {2018})}\BibitemShut {NoStop}%
\end{thebibliography}%

\end{document}